# Density of states and zero Landau level probed through capacitance of graphene

L. A. Ponomarenko[1], R. Yang[1], R. V. Gorbachev[1], P. Blake[1], M. I. Katsnelson[2], K. S. Novoselov[1], A. K. Geim[1]

[1]*Manchester Centre for Mesoscience & Nanotechnology,*
*University of Manchester, Manchester M13 9PL, United Kingdom*
[2]*Theory of Condensed Matter, Institute for Molecules and Materials, Radboud University*
*Nijmegen, Heyendaalseweg 135, 6525 AJ Nijmegen, The Netherlands*

We report capacitors in which a finite electronic compressibility of graphene dominates the electrostatics, resulting in pronounced changes in capacitance as a function of magnetic field and carrier concentration. The capacitance measurements have allowed us to accurately map the density of states $D$, and compare it against theoretical predictions. Landau oscillations in $D$ are robust and zero Landau level (LL) can easily be seen at room temperature in moderate fields. The broadening of LLs is strongly affected by charge inhomogeneity that leads to zero LL being broader than other levels.

PACS numbers: 73.22.Pr, 81.05.ue

One of the most celebrated consequences of the Dirac-like electronic spectrum of graphene is its zero LL centered at the neutrality point (NP) and shared by hole- and electron- like carriers [1]. Although the electronic properties become particularly interesting near the NP, it has proven difficult to probe this regime by transport measurements. The problem is not only the potential fluctuations that move the Dirac point spatially and average out interesting features [2]. The situation is additionally tangled because transport coefficients become non-monotonic at the NP and sensitive to scattering details, even in high magnetic fields $B$ [3]. Capacitance measurements provide an alternative. If graphene is incorporated in a capacitor as one of its electrodes, there appears a significant contribution into the total capacitance $C$ due to the electronic compressibility. This contribution is often referred to as quantum capacitance $C_q = e^2 D$ and is a direct measure of the density of state $D(E) = dn/dE$ at the Fermi energy $E_F$ ($e$ is the electron charge; $n$ the carrier concentration) [4,5].

As for experimental studies of $C_q$, graphene is unique for two reasons. First, it has an atomically thin body, which allows capacitors in which the classical, geometrical contribution plays a minor role so that $C_q$ can completely dominate the device's electrostatics. Second, $D$ is a strong function of $E_F$ and, therefore, $C_q$ can be changed by applying gate voltage $V_g$. This distinguishes graphene from conventional two-dimensional systems in which $C_q$ is usually a small and constant contribution that is difficult to discern experimentally [5]. Thanks to the $V_g$ dependence, several groups have already reported the observation of $C_q$ of graphene [6-9]. Their measurements showed the expected V-shape dependence centered at the NP. However, the electron and hole branches were often strongly asymmetric [6,7], contrary to expectations, and the absolute value of $C_q$ was either impossible to determine [8,9] or it disagreed with theory [6]. Most recently, capacitance measurements were also employed to prove the gap opening in double-gated bilayer graphene [10].

In this work we report large (~100x100 μm$^2$) graphene capacitors with a thin (≈10 nm) dielectric layer and a high carrier mobility μ of ≈10,000 cm$^2$/Vs maintained after the fabrication. In our devices, $C_q$ is no longer merely a correction but reaches ≈30% of the measured capacitance, so that its changes with varying $V_g$ and $B$ are very pronounced. This allowed us to compare the quantum capacitance against theoretical predictions, the task proven impossible in the previous studies. Second, we show that, in capacitance experiments, zero LL is extremely robust with respect to temperature $T$ and disorder and clearly seen at room $T$ in fields of 10 T (note that $B$ as high as 30T were required to observe the quantum Hall effect at room $T$ [11]). Third, we have analyzed the broadening of LLs as a function of $B$ and $T$, and found that a charge inhomogeneity (electron-hole puddles) is as important as scattering in defining LLs' width. In particular, the inhomogeneity results in zero LL being wider than other LLs, which is directly visible from our experimental curves. This observation seems in contradiction to the suggestion that zero LL is exceptionally narrow [12]. Fourth, we observe no splitting at the NP in fields up to 16 T, reported in transport experiments for devices of similar quality [13].

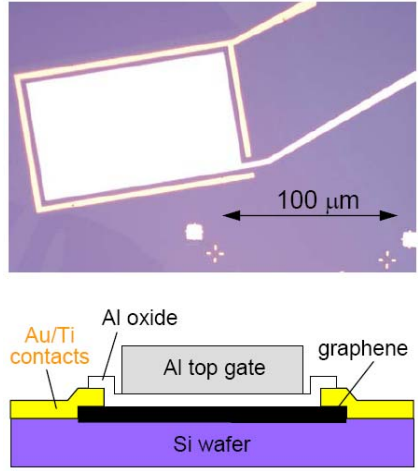

FIG. 1 (color online). Photograph of one of our devices (top) and its schematics (bottom). Graphene is connected to the measurement circuit by a contact deposited along the device's perimeter. The second capacitor plate is an Al electrode (central rectangular in the top panel).

We developed the following technology for making graphene capacitors (Fig. 1). Large crystals were obtained by mechanical cleavage and deposited on top of an oxidized Si wafer. High-resistivity (>1 k$\Omega$·cm) wafers were essential to avoid a contribution from parasitic capacitances. 1 nm of Al was then deposited on top of graphene in the presence of oxygen ($O_2$ pressure of ~0.1 mBar), which resulted in an initial layer of aluminum oxide. A thick (100 nm) layer of Al was then evaporated on top of graphene in high vacuum, forming the second plate of a flat capacitor. Although only 1nm of Al oxide was deposited, the resulting gate dielectric appeared to be much thicker (~10 nm) as found by atomic force and electron microscopy. This is attributed to the oxidation of the Al gate from below by air that diffused through pores in the initial oxide. Importantly, the technology allowed us to avoid any noticeable reduction in $\mu$ due to the gate dielectric and change $E_F$ between ±0.5eV. $\mu$ was measured by using multi-terminal Hall bar devices fabricated in parallel with the capacitors. The Hall bars had some areas covered with the top gate and some left open. The measured $\mu$ were statistically indistinguishable for the two areas. The only notable difference was that, in the top-gated devices, the NP was always close to zero $V_g$ (remnant doping $n_D < 10^{11}$ cm$^{-2}$). In contrast, non-gated graphene typically exhibited $n_D \geq 10^{12}$ cm$^{-2}$. We attribute this behavior to screening out of the average electrostatic potential by the nearby Al gate.

A typical dependence of $C$ on $V_g$ applied between graphene and the Al gate is plotted in Fig. 2. We have found little contribution from parasitic capacitances (<1pF) and, to illustrate this fact, Fig. 2a also shows $C$ measured in an alternative approach (circles) by studying magnetocapacitance oscillations (figure caption). $C_q$ contributes as a series capacitor [4,5]: $1/C = 1/C_{ox} + 1/C_q$ where $C_{ox}$ is the gate independent geometrical capacitance of the oxide layer. The measured dependence in Fig. 2a is accurately described by two parameters [4-10,14]: 1) $C_{ox}$ that gives the saturation value in the limit of high $n$ (where $D$ becomes large so that $C_q >> C_{ox}$) and 2) the Fermi velocity $v_F$ in graphene that describes the rate of changes in $C$ at low $n$. The dashed curve in Fig. 2a shows the best fit that yields $C_{ox} \approx 0.47$ μF/cm² and $v_F \approx 1.15 \pm 0.1 \cdot 10^6$ m/s (the error indicates the reproducibility for four different devices). The found $v_F$ is in excellent agreement with other experiments (e.g., [15,16]) and theory [1,17]. The $C_{ox}$ value is the only real fitting parameter and corresponds to the classical capacitance expected for our oxide thickness of ~10nm.

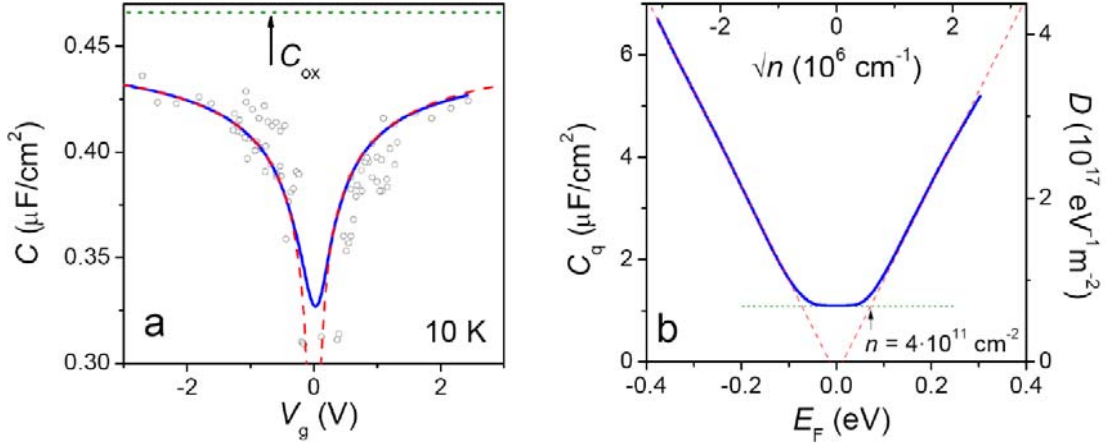

FIG. 2 (color online). Graphene capacitance in zero $B$. (a) Capacitance (per unit area) of the device in Fig. 1 (solid curve). The dashed curve is the best fit and the horizontal line shows $C_{ox}$. Open symbols are the alternative measurements using the periodicity of magnetocapacitance oscillations. For a given $B$, each period in gate voltage $\Delta V_g$ corresponds to the filling of one LL, and this requires an increase in carrier concentration $\Delta n = 4eB/h$ where $h$ is the Planck constant. $C$ is then calculated as $e\Delta n/\Delta V_g$. Despite the lower accuracy of this approach (due to oscillations in $C_q$), it provides a valuable crosscheck. (b) $C_q$ as function of $E_F$ (bottom axis) and $n$ (top axis). The right axis plots the density of states, $D = C_q/e^2$. The extrapolation lines (dashed) in Fig. 2b should cross at zero, and the small vertical offset can be explained by parasitic capacitances that we deliberately did not include in the analysis to minimize the number of fitting parameters.

By integrating the experimental curve in Fig. 2a, we can find $n$ for a given $V_g$ and replot the data as a function of $n$ and $E_F = \hbar v_F \sqrt{\pi |n|}$ (note that the standard equation $n \propto V_g$ no longer holds for thin gate dielectrics). The resulting curve in Fig. 2b is a textbook behavior for the quantum capacitance of Dirac fermions. As expected, $D(E_F)$ exhibits equal (within 3%) slopes for the valence and conduction bands, which are given by the single material parameter $v_F$. The prominent flat region at the NP is attributed to electron-hole puddles [16] that smear the sharp dip in $C$ at the NP. The smearing in Fig. 2b covers a region of $\delta n \approx \pm 4 \cdot 10^{11}$ cm⁻² which is close to $\delta n$ typically observed in transport experiments [2]. By comparing Figs. 2a&b, one can see that electron-hole puddles dominate the spectral smearing near the NP (because the energy smearing $\delta E \propto \delta n/E_F$).

In perpendicular $B$, our capacitors exhibit pronounced oscillations (Fig. 3a). Zero LL is centered at the NP (zero $E_F$) as expected for graphene's electronic spectrum [1,15]. First, let us understand the overall shape of the $C_q$ curves. $D(E)$ can be considered as a superposition of LLs having the Lorentzian shape [18]. Fig. 3b shows results of our numerical analysis. For graphene with LLs equally broadened by scattering and, therefore, having the same width $\Gamma$, we could not reach any satisfactory agreement with our experiment for any $\Gamma$. Theory predicts stronger oscillations in $C_q$ (dashed curve) whereas in reality the peaks become progressively smaller at small $n$. To explain the latter, we take into account the charge inhomogeneity inferred from Fig. 2b. This provides a natural explanation for zero LL being broader (in terms of $\delta E$) than other LLs and, thus, having the smallest height in Fig. 3a. It also allows us to account for the overall shape of the $C_q(V_g)$ curve. Furthermore, we have often observed an electron-hole asymmetry in $C_q$ such that the oscillations for electrons are more pronounced than those for holes (Fig. 3a). The degree of this asymmetry varied from sample to sample, and we attribute it to an asymmetric potential landscape. To this end, we have modeled the charge inhomogeneity as a sum of two Gaussians, which can be due to, for example, macroscopic regions with different chemical doping. The resulting dependence (solid curve) reproduces the essential features of the observed asymmetry. In general, the analysis in Fig. 3b shows that, at low $T$, the spectral broadening is mostly due to electron-hole puddles and the finite $\mu$ plays little role for zero LL.

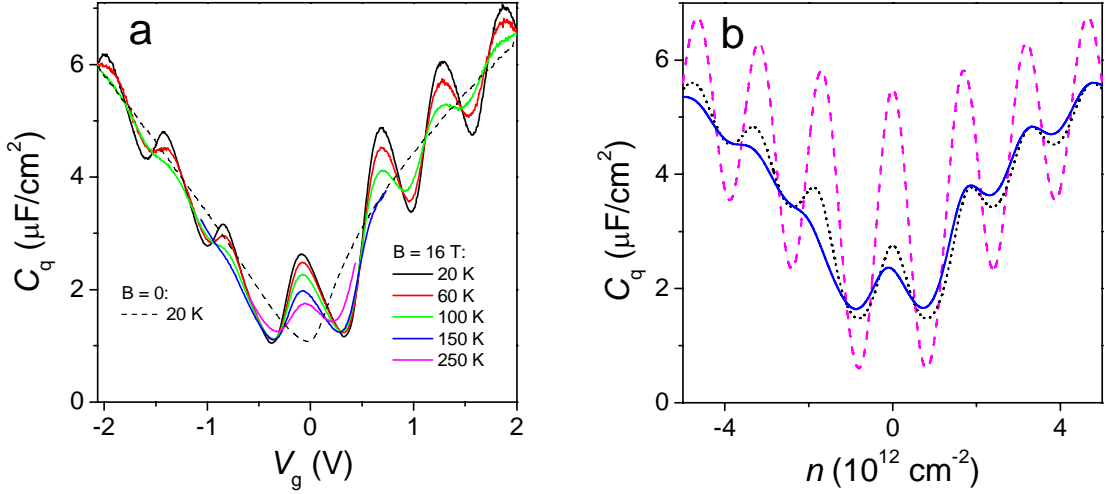

FIG. 3. (colour online) Magnetic oscillations in the density of state. (a) Experiment: $C_q = e^2 D$ as a function of $V_g$ at various $T$. (b) Theory: $C_q$ in 16 T at low $T$. The dashed curve shows the oscillations if all LLs were equally broadened with the half width at half maximum (HWHM), $\Gamma =15$meV [19]. The dotted curve takes into account electron-hole puddles and was obtained as a convolution of the above broadening with an added inhomogeneity in $n$ which was modeled by a Gaussian distribution [20] with standard deviation $\delta n = 4 \times 10^{11}$ cm$^{-2}$ (the value found in Fig. 2). The solid line is a result of a similar convolution but assuming an asymmetric doping, which was modeled as a sum of two Gaussians with height ratios of 7:3; $\delta n =3.5 \times 10^{11}$ cm$^{-2}$ for both; and the separation of $7 \times 10^{11}$ cm$^{-2}$.

The $T$ dependence in Fig. 3a has allowed us to find cyclotron mass $m_c$ at different $V_g$. The analysis is the same as for the case of Shubnikov-de Haas oscillations [15], and our results (not shown) are essentially identical, yielding $m_c \propto n^{1/2}$ with a single fitting parameter $v_F \approx 1.1 \pm 0.1 \cdot 10^6$ m/s. Note that such an analysis is not applicable at the Dirac point where the nominal value of $m_c$ is zero. This emphasizes differences between transport and capacitance measurements. The latter allow us to directly assess $D(E)$ whereas interpretation of transport experiments is generally more complicated and sensitive to scattering details. Also, note that under the same conditions our reference multi-terminal devices exhibited clear quantum Hall effect plateaus and wide zeros in resistance (not shown; for examples, see Refs. [1,15]). This difference can be explained by the fact that the capacitance spectroscopy probes both extended and localized states, whereas electron transport occurs over the former.

Because zero LL is hard to assess in transport experiments, we focus below on this particular feature. Fig. 4 details its behavior with changing $T$ and $B$. We start with pointing out two qualitative observations for zero LL. First, we do not see any hint for zero LL's splitting in fields up to 16T and $T$ down to 1K [13]. We believe that, if present, this splitting for the $D(E)$ curves should be smeared by electron-hole puddles. Second, zero LL can be clearly seen in $C_q$ even at room $T$ in fields down to 10T and an increase in $D_{NP} \equiv D(0)$ can be detected in $B$ as low as 5T (Figs. 3a&4a). To analyze the observed $T$ dependence, Fig. 4b plots $D_{NP}$ as a function of $T$. As for theory, each of thermally broadened LLs contributes to $D_{NP}$ as [22]

$$D_i(T) = \frac{\Delta n}{\pi} \int_0^\infty e^{-\Gamma_0 t} \cos(E_i t) \frac{\pi T t}{\sinh(\pi T t)} dt$$

where $\Gamma_0$ is the HWHM at $T=0$ and $E_i = \pm v_F \sqrt{2\hbar eB|i|}$ is the energy of $i$-th LL ($i = 0, \pm 1, \ldots$). The dashed theory curve in Fig. 4b plots $D_{NP}$ if only zero LL is taken into account, which is justified because at 16 T the distance to the next LL is larger than $\Gamma_0$ and $T$. If the next LL ($i = \pm 1$) is included, this does not improve the fit (solid curve). We find $\Gamma_0 \approx 30$ meV that is twice larger than the one used in the analysis in Fig. 3b because it now includes the smearing $\delta E$ due to charge inhomogeneity. The inset in Fig. 4b replots the data from the main figure in terms of the LL broadening by using the standard expression $\Gamma_\Sigma = \Delta n / \pi D_{NP}$ where $\Delta n$ is the number of carriers at each LL. This presentation is more intuitive because one can readily see that zero LL have an intrinsic broadening $\Gamma_0$ and broadens approximately linearly with $T$ as $\Gamma_\Sigma \approx \Gamma_0 + \beta T$ where $\beta \approx 1.1 \pm 0.2$, in agreement with general expectations.

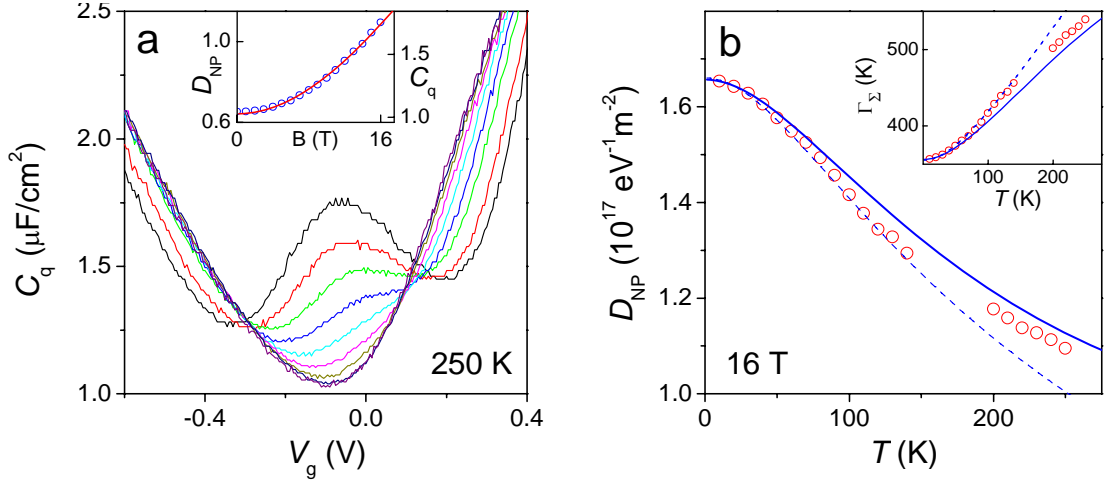

FIG. 4. (color online) Density of states near the NP. (a) The emergence of zero LL with increasing $B$ near room $T$. The curves cover the range from 0 to 16 T in steps of 2T. Zero LL becomes clearly visible above 10 T. We avoided $T$ above 250K because the measurements become hysteretic due to an increased mobility of adsorbates, similar to transport experiments [21]. The inset shows $C_q$ and $D$ at the NP as a function of $B$ in units of $\mu F/cm^2$ and $10^{17}$ eV$^{-1}$m$^{-2}$, respectively. The solid curve is the best fit [22]. (b) $T$ dependence of $D_{NP}$ at 16 T (symbols). The curves are theory fits. Inset: same as in the main figure but in terms of the width of zero LL.

Our model can also describe the field dependence of $D$. The inset in Fig. 4a shows $D_{NP}$ as a function of $B$ at 250K, and the solid curve is the theory fit [22] using $\Gamma_\Sigma \approx 50$meV found in Fig. 4b. In general, our analysis in Figs. 3 and 4 shows that the measured $D(E)$ and its quantization are in good agreement with theory.

In summary, changes in the density of states can completely dominate the behavior of graphene capacitors with a thin dielectric layer. This provides a useful approach for studies of quantization phenomena in graphene and, especially, its lowest Landau level that is difficult to assess in transport experiments. In the capacitance measurements, zero LL is easily observable at room $T$. This requires $B$ of only 10T, which is probably the record for the Landau quantization observed in any material. However, contrary to previous experiments using similar-quality devices, we find no indications that zero LL is narrower than other LLs [12] and no splitting of zero LL is observed in fields up to 16T [13]. This can be due to the difference in contributions from extended and localized states in transport and capacitance measurements but may also call for alternative interpretations. Finally, we mention that our devices prove the feasibility of bipolar variable capacitors that can have large $C \sim 1\mu F/cm^2$ and be controlled by small voltages (~0.1V).

EPAPS Document for
*Density of states and zero Landau level probed through capacitance of graphene* by Ponomarenko *et al*.

The quantum capacitance is proportional to the thermally smeared density of states $D(E)$ and is given by

$$C = e^2 \int dE D(E) \left[ -\frac{\partial f(E + eV_g)}{\partial E} \right]$$

The Landau levels are $E_j = \hbar\omega_c \sqrt{j}$, $\hbar\omega_c = \frac{\sqrt{2}\hbar v_F}{l_B}$ is the cyclotron quantum, $l_B = \sqrt{\frac{\hbar}{eB}}$ is the magnetic length.

Taking into account the broadening of LL and assuming that it is Lorenzian, we obtain

$$D(E) = \frac{2}{\pi l_B^2} \left\{ \frac{\Gamma_0}{\pi} \frac{1}{E^2 + \Gamma_0^2} + \sum_{n=1}^{\infty} \left[ \frac{\Gamma_j}{\pi} \frac{1}{(E-E_j)^2 + \Gamma_j^2} + \frac{\Gamma_j}{\pi} \frac{1}{(E+E_j)^2 + \Gamma_j^2} \right] \exp(-\alpha E_j) \right\}$$

per unit area. Here we introduce an energy cutoff $1/\alpha$ of the order of the bandwidth. Otherwise, the sum logarithmically diverges for large $j$.

It is convenient to use the following representation $\frac{\Gamma}{\pi} \frac{1}{(E-z)^2 + \Gamma^2} = \text{Re} \frac{1}{\pi} \int_0^{\infty} dt \exp[-\Gamma t + i(E-z)t]$ and the integral $-\int_{-\infty}^{\infty} dE e^{iEt} \frac{\partial f(E-z)}{\partial E} = e^{izt} \frac{\pi T t}{\sinh(\pi T t)}$.

Thus, we obtain

$$C = \frac{4e^2}{\pi l_B^2} \left\{ \Lambda_0(eV_a) + \sum_{j=1}^{\infty} \left[ \Lambda_j(eV_a - \hbar\omega_c\sqrt{j}) + \Lambda_j(eV_a + \hbar\omega_c\sqrt{j}) \right] \exp(-\alpha E_j) \right\},$$

$$\Lambda_j(z) = \frac{1}{\pi} \int_0^{\infty} dt e^{-\Gamma_j t} \cos(zt) \frac{\pi T t}{\sinh(\pi T t)}$$

which is the equation used in the main article. The integral over $t$ is rapidly converging which makes this presentation convenient for numerical calculations.

Our model can also describe the field dependence of $D(E)$. However, to cover the whole range of $B$, we could not continue using the Lorentzian broadening. This is because at low B separation between LLs is small and results in the known divergence for large $j$ as more and more LLs are taken into account.

Instead, we use the Gaussian broadening of LLs, which leads to

$$D(E) = \frac{2}{\pi l_B^2} \left\{ \frac{1}{\sqrt{2\pi}\sigma_0} \exp\left(-\frac{E^2}{2\sigma_0^2}\right) + \sum_{j=1}^{\infty} \frac{1}{\sqrt{2\pi}\sigma_j} \left[ \exp\left(-\frac{(E - \hbar\omega_c\sqrt{j})^2}{2\sigma_j^2}\right) + \exp\left(-\frac{(E + \hbar\omega_c\sqrt{j})^2}{2\sigma_j^2}\right) \right] \right\} \text{ where}$$

$\sigma_j$ is the standard deviation in the Gaussian distribution (parameter of the order of half width). Assuming $\sigma_j = \sigma$ (that is, independent of $j$) one can calculate the sum at the NP $D(E=0) \equiv D_{NP}$.

Taking into account that $1 + 2\sum_{j=1}^{\infty} \exp(-2\gamma j) = 1/\tanh(\gamma)$, we obtain

$$D_{NP}(B) = \left(\frac{2}{\pi}\right)^{3/2} \frac{\sigma}{(\hbar v_F)^2} \frac{\gamma}{\tanh \gamma}$$

where parameter $\gamma = (\hbar\omega_c / 2\sigma)^2$ depends on the ratio of the cyclotron energy $\hbar\omega_c = v_F\sqrt{2\hbar eB}$ and the LL width. The function obtained in the last equation was used to fit the experimental data presented in the inset of Fig. 4a using $\sigma$ as the only fitting parameter ($\sigma$ = 60 meV). Notice, that this equation applies for any magnetic field including $B = 0$, in which case $\frac{\gamma}{\tanh \gamma} = 1$.